\documentclass[conference]{IEEEtran}

\usepackage{amssymb}
\usepackage{amsmath}
\usepackage{amsfonts}
\usepackage{dsfont} 
\usepackage[utf8]{inputenc}
\usepackage{amsthm}
\usepackage{cite}
\usepackage{subfigure}
\usepackage{color}
\usepackage{times}
\usepackage{graphicx}
\usepackage{psfrag}
\usepackage[dvipsnames]{xcolor}

\makeatletter
\newcommand*\rel@kern[1]{\kern#1\dimexpr\macc@kerna}
\newcommand*\widebar[1]{%
  \begingroup
  \def\mathaccent##1##2{%
    \rel@kern{0.8}%
    \overline{\rel@kern{-0.8}\macc@nucleus\rel@kern{0.2}}%
    \rel@kern{-0.2}%
  }%
  \macc@depth\@ne
  \let\math@bgroup\@empty \let\math@egroup\macc@set@skewchar
  \mathsurround\z@ \frozen@everymath{\mathgroup\macc@group\relax}%
  \macc@set@skewchar\relax
  \let\mathaccentV\macc@nested@a
  \macc@nested@a\relax111{#1}%
  \endgroup
}
\makeatother

\newcommand{\bm}{\mathbf}
\newcommand{\be}{\begin{equation}}
\newcommand{\ee}{\end{equation}}
\newcommand{\bea}{\begin{eqnarray}}
\newcommand{\eea}{\end{eqnarray}}

\newcommand{\bA}{{\bm A}}
\newcommand{\bI}{{\bm I}}

\newcommand{\bD}{{\bf D}}

\newcommand{\bP}{{\bf P}}

\newcommand{\bH}{{\bf H}}

\newcommand{\bd}{{\bf d}}
\newcommand{\bs}{{\bf s}}

\IEEEoverridecommandlockouts
\begin{document}

\title{Downlink Precoding for FBMC-based Massive MIMO with Imperfect Channel Reciprocity}

\author{\normalsize Hamed Hosseiny$^\dagger$, Arman Farhang$^*$ and Behrouz Farhang-Boroujeny$^\dagger$  
\\$^\dagger$ECE Department, University of Utah, USA, \\
$^*$Department of Electronic and Electrical Engineering, Trinity College Dublin, Ireland.  \\
Email: \{hamed.hosseiny, farhang\}@utah.edu, \{arman.farhang\}@tcd.ie}
\maketitle

\begin{abstract}

In this paper, a practical precoding method for the downlink of filter bank multicarrier-based (FBMC-based) massive multiple-input multiple-output (MIMO) is developed. The proposed method includes a two-stage precoder consisting of a fractionally spaced prefilter (FSP) per subcarrier for flattening/equalizing the channel across the subcarrier band, followed by a conventional precoder whose goal is to concentrate the signals of different users at their spatial locations. This way, each user receives only the intended information. In this paper, we take note that channel reciprocity may not hold perfectly in practical scenarios due to the mismatch of radio chains in uplink and downlink. Additionally, channel state information (CSI) at the base station may not be perfectly known. This, together with imperfect channel reciprocity can lead to detrimental effects on the downlink precoder performance. We theoretically analyze the performance of the proposed precoder in the presence of imperfect CSI and channel reciprocity calibration errors. This leads to an effective method for compensating these effects. Finally, we numerically evaluate the performance of the proposed precoder. Our results show that the proposed precoder leads to an excellent performance when benchmarked against OFDM.

\end{abstract}
\begin{IEEEkeywords}
FBMC, multiuser, precoder, massive MIMO, downlink.
\end{IEEEkeywords}

\vspace{-4mm}
\section{Introduction}
The success of massive multiple input multiple output (MIMO) technology in the recent roll-out of the fifth generation wireless systems (5G) is an advocate on the importance of multiple antenna techniques for future networks \cite{giordani2020toward}. Thus, massive MIMO will be among the key building blocks that underpin the future of 5G Advanced and the sixth generation wireless networks (6G) \cite{saad2019vision}. The shortcomings of the orthogonal frequency division multiplexing (OFDM) such as its high sensitivity to synchronization errors were taken on board in the design of 5G new radio (5G NR) standard by the introduction of the flexible subcarrier spacings \cite{5gsystem}. However, OFDM still suffers from bandwidth efficiency loss considering the extended cyclic prefix of the length $25\%$ of symbol duration, \cite{5gsystem}. Furthermore, the advent of new applications, such as autonomous driving, where wireless channels become highly time varying, call for alternative waveforms that are more resilient than OFDM to the time variations of the channel \cite{schulz2017latency}. Filter bank multicarrier (FBMC) is one of a kind with a high bandwidth efficiency and resilience to the synchronization errors and the channel time variations \cite{farhang2011ofdm,nissel2017filter, 7369549}.

The above observations, clearly, justify the significance of exploring FBMC-based massive MIMO as a candidate technology for the future wireless systems. FBMC-based massive MIMO was first introduced in \cite{farhang2014}. In this work, the authors showed how FBMC benefits from the channel flattening effect of massive MIMO to widen the subcarrier bands and thus further improve the bandwidth efficiency. As a follow up contribution, in \cite{farhang2014pilot}, the authors addressed the pilot contamination problem in the uplink of FBMC-based Massive MIMO systems. Further studies in \cite{rottenberg2018performance, singh2019uplink} provide the mean squared error (MSE) and sum-rate performance of FBMC in the uplink of massive MIMO channels, respectively. Channel estimation and equalization aspects of FBMC-based massive MIMO were covered in \cite{hosseiny2021fbmc,9508836} and \cite{aminjavaheri2015frequency,aminjavaheri2018filter}, respectively. While ideal scenarios are considered in a large body of the available literature on the topic, in a more recent work, we focused on the practical aspects of FBMC-based massive MIMO systems \cite{hosseiny2021fbmc}. In particular, we investigated imperfect channel state information (CSI) effects on the performance of FBMC-based massive MIMO in both co-located and distributed antenna setups \cite{hosseiny2021fbmc}. 

As of today, most publications on FBMC-based massive MIMO have focused on the uplink, \cite{aminjavaheri2018filter,aminjavaheri2017prototype,Hoss2006:Spectrally,hosseiny2021fbmc,singh2019time}. A few of these publications have assumed perfect reciprocity and, accordingly, have noted that the proposed uplink detection methods may be reversed to design precoders for the downlink of the same link, e.g., see \cite{aminjavaheri2018filter}.  However, the assumption of perfect reciprocity may not be valid, both due to channel aging in time division duplexing (TDD) and the differences in radio chains (even after calibration) in the uplink and downlink directions.  Works such as \cite{chopra2020blind,mi2017massive} have investigated imperfect reciprocity problem for narrow-band systems. While the extension of such analysis to OFDM-based systems is straightforward, it requires particular attention and investigation to be extended to FBMC-based massive MIMO.

In this paper, we focus on the downlink precoder design for FBMC-based massive MIMO systems, in presence of channel estimation and reciprocity calibration errors. To this end, we first formulate the downlink transmission for FBMC-based massive MIMO systems, assuming a perfect reciprocal channel. Following the uplink detector structure developed in \cite{hosseiny2021fbmc}, we propose a two-stage precoder structure. The first stage of this structure involves a short per-user fractionally spaced prefilter (FSP) (equivalent to a fractionally spaced equalizer (FSE)) at each subcarrier for flattening the equivalent channel over the respective band. At the second stage, any of the conventional linear precoders based on the maximum ratio transmission (MRT), the zero forcing (ZF), or the minimum mean square error (MMSE) techniques can be deployed. Next, taking into account the imperfect CSI and reciprocity calibration error, we analytically derive their effect on the received signals at the user terminals. We find that the imperfect CSI and calibration error effects converge to a fixed, error statistics dependent scaling factor, which is the same for all the subcarriers. 

The rest of the paper is organized as follows. Section~\ref{sec:flatfbmc} presents principles of FBMC in the downlink of massive MIMO assuming a flat subcarrier followed by Section~\ref{sec:selective_precoding} proposing two-stage precoding to overcome frequency selectivity in the channel. In Section~\ref{sec:calibration} we extend our analysis to the imperfect CSI and channel reciprocity by modeling the reciprocity error and proposing compensation methods to relax the effect of both errors. Section \ref{sec:simulation} provides numerical analysis, confirming the validity of our claims through simulations. Finally, the paper is concluded in Section \ref{sec:conclusion}.

\vspace{3mm}
\noindent\textit{Notations:} Matrices, vectors and scalar quantities are denoted by boldface uppercase, boldface lowercase and normal letters, respectively.  $A(m,l)$ represents the element in the $m^{\rm th}$
row and the $l^{\rm th}$ column of $\bA$ and $\bA ^{-1}$ signifies the inverse of $\bA$. $\bI_M$ is the identity matrix of size $M \times M$. Superscripts $(\cdot)^{-1}$, $(\cdot)^{\rm T}$, $(\cdot)^{\rm H}$ and $(\cdot)^*$ indicate inverse, transpose, conjugate transpose, and conjugate operations, respectively. $\mathfrak{R}\{\cdot\}$,  $\mathds{E}\{\cdot\}$, ($\downarrow M$) and $\star$  represent real value, expectation, $M$ fold decimation and linear convolution operators, respectively. 
Finally, $\delta_{ij}$ represents the Kronecker delta function.

 \vspace{-1mm}
\section{Downlink FBMC System Model}
\label{sec:flatfbmc}
In FBMC, real-valued data symbols are placed on a regular time-frequency grid with the time and frequency spacings of $T/2$ and $1/T$, respectively. Each data symbol on the grid has a $\pm\frac{\pi}{2}$ phase difference with its neighbours. This is to avoid interference between the data symbols and hence make them orthogonal in the real domain. The data symbols are pulse-shaped with a prototype filter, $f[l]$ where $f[l]$ is designed such that $q[l]=f[l]\star f^*[-l]$ satisfies the Nyquist criterion. Therefore, assuming $M$ number of subcarriers, the Nyquist pulse $q[l]$ has zero crossings every $M$ samples. Considering a narrowband be subcarrier such that the data symbols experience approximately a  flat fading, similar to OFDM, per subcarrier precoding can be deployed in the downlink \cite{marzetta2016fundamentals}.

Let us consider a single-cell massive MIMO setup including a BS equipped with $N$ antennas and $K$ single-antenna users.  Let $d^k_{m,n}$ be the real-valued data symbol of user $k$ at the frequency index $m$ and the time index $n$. For each frequency-time instant $(m,n)$, the  precoder collates  data symbols  $d_{m,n}^k, k=1,2,\cdots,K$ from all users and constructs the transmit signal vector 
\be 
\bs_{m,n} = \bP_m^{\rm H} \bd_{m,n},\label{eq:s_{m,n}}
\ee
where $\bP_m$ is the $K \times N$ precoding matrix that can be chosen from any of the common linear precoders, namely, MRT, ZF, and MMSE, i.e.,
 \be
 \bP_m = \begin{cases}
\bD_m^{-1} \bH_m, & \text{for MRT,}\\
 \big(\bH_m \bH_m^{\rm H}\big)^{-1} \bH_m, & \text{for ZF,}\\
 \big(\bH_m \bH_m^{\rm H}+\sigma^2_{\eta} \bI_K\big)^{-1}\bH_m, & \text{for MMSE}.
\end{cases}
\label{eq:combiners}
\ee
Here, $\bH_m$ is the $K\times N$ channel matrix at the center of  subcarrier $m$ with elements $H_m(k,i)$ representing the channel gains between user $k$ and BS antenna $i$, i.e.,  $H_m({k,i}) \triangleq \sum_{l=0}^{L-1} h_{k,i}[l] e^{-j\frac{2 \pi m l}{M}}$, where $h_{k,i}[l]$ is the respective channel impulse response. In MRT, the $K \times K $ diagonal matrix $\bD_m$ normalizes the precoder outputs with the coefficients $D_m^{k,k} = \sum _{i=0}^{N-1}|H_m({k,i})|^2$. Assuming reciprocal channels in the uplink and downlink, the estimated channel responses in the uplink phase are used in the above precoders. Note that in the reciprocal scenario, downlink channel matrix is the transposed version of the uplink channel matrix. 

After precoding, the transmit signal at the BS antenna $i$ can be formed by passing the symbols $s^i_{m,n}$ through the synthesis filterbank (SFB)
\begin{equation}
x_i[l] = \sum_{m=0}^{M-1} \sum_{n=-\infty}^{ \infty} s^i_{m,n} f_{m,n}[l] ,
\label{bbeq}
\end{equation}
where
$f_{m,n}[l] = f\big[l-n\frac{M}{2}\big] e^{j2\pi ml/M}e^{j\pi (m+n)/2}$
is the modulated, time shifted, and phase-adjusted pulse-shape that carries $s^i_{m,n}$. Finally, the received signal at user $k$ can be obtained as
\be
r_k[l] = \sum_{i=0}^{N-1} x_i[l] \star h_{k,i}[l] + \eta_k[l],
\label{eq:rec_antenna}
\ee
where $\eta_k[l] \sim \mathcal{CN}(0,\sigma^2_{\eta})$ is the additive white Gaussian channel noise at the terminal $k$ with the variance $\sigma^2_{\eta}$. 
Considering co-located BS antennas, we can assume the same PDP between the BS antennas and any given user $k$. This PDP is denoted by  $p_{k}[l]$ for $l=0,\ldots,L-1$, where $L$ is the channel length. Accordingly, for user $k$, the channel taps are independent of one another and their distribution follows $\mathcal{CN}(0,p_{k}[l])$. \textcolor{black}{Furthermore, we assume the average transmit power of unity for each user terminal}

Considering perfect synchronization and channel knowledge at the BS, the data symbols of each user can be extracted as
\be
\hat{d}_{m,n}^k = \Re \{(r_k[l] \star \color{black} f_{m,n}[l]\color{black})|_{{l=\frac{nM}{2}}} \}.
\label{d^hat}
\ee
In FBMC, the assumption of flat fading subcarrier channels can never be satisfied no matter how narrow the subcarrier bands are made. Beside, for the purpose of spectral efficiency, it is always desirable to keep the subcarrier bands as wide as possible; \textcolor{black}{see \cite{farhang2014} and \cite{farhang2014massive} } for some explanations along this line. Moreover, we may recall from \cite{farhang2014} that channel hardening effect in massive MIMO systems allows one to widen the subcarrier bands. 
However, investigations in \cite{aminjavaheri2018filter} has revealed that  this hardening effect flattens the channel to a limit. Hence, an additional equalization/precoding step is required for (near) perfect flattening of the channel over each subcarrier band. Moreover, imperfections in the available CSI that may originate from channel estimation error and aging in TDD, as well as reciprocity calibration errors have to be compensated for. To tackle these practical issues, in Section~\ref{sec:selective_precoding}, we propose a prefilter for flattening each subcarrier channel, and in  Section~\ref{sec:calibration}, we address the problems of imperfect CSI and channel reciprocity.

\section{Subcarrier Prefiltering} \label{sec:selective_precoding}
By expanding \eqref{d^hat} and using \eqref{eq:rec_antenna}, we obtain
\begin{align}\label{eq:dhatmnk1}
\hat{d}_{m,n}^k \!= &\Re \bigg\{ \!\sum_{i=0}^{N-1} \! \big(x_i[l] \star h_{k,i}[l] \star \color{black}f_{m,n}[l]\color{black}\big)\big|_{l=\frac{nM}{2}} + \eta_{m,n}^k \bigg\},
\end{align}
where $\eta_{m,n}^k$ is the noise effect after filtering and phase adjustment. Recalling \eqref{eq:s_{m,n}} and \eqref{bbeq}, \eqref{eq:dhatmnk1} can be simplified as
\begin{align}
&\hat{d}_{m,n}^k =   \Re \bigg\{ \sum_{m',n',k'}   d_{m',n'}^{k'} g_{m,m'}^{k,k'}[n-n']  +\eta_{m,n}^k \bigg\},
\label{eq:eqxpandedEstimate}
\end{align}
where
\begin{align}\label{eq:g_{mm'}^{k,k'}[n]}
 g_{mm'}^{k,k'}[n] = \big(f_{m'}[l] \star h_{k,k',m}^{\rm (eqvlt)}[l] \star f_m^*[l]  \big)\big|_{l=\frac{nM}{2}},
\end{align}
\be\label{eq:equivalent-channel-general}
h_{k,k',m}^{\rm (eqvlt)}[l]=\sum_{i=0}^{N-1}(P_m^{k',i})^*h_{i,k}[l],
\ee
and the coefficients $P_m^{k,i}$ are the precoder coefficients given by \eqref{eq:combiners}. As shown in \cite{ngo2013energy}, for a large number of antennas, all the above precoders converge to $\frac{1}{N} \bH_m$.
 Accordingly, the equivalent channel between the user terminal $k$ and the precoder input intended for the $k'$th user over the subcarrier band $m$ may be expressed as 
\be\label{eq:equivalent-channel}
h_{k,k',m}^{\rm (eqvlt)}[l]=\frac{1}{N}\sum_{i=0}^{N-1}(H_m({k',i}))^*h_{i,k}[l].
\ee
Using the law of large numbers, for a large number of antennas at the BS,  $h_{k,k',m}^{\rm (eqvlt)}[l]$ vanishes to zero, when $k\ne k'$. Additionally, when $k=k'$, it can be shown that $h_{k,k,m}^{ \rm (eqvlt)}[l]$ in \eqref{eq:g_{mm'}^{k,k'}[n]} converges to \cite{aminjavaheri2018filter,hosseiny2021fbmc}
\be\label{eq:pmk[l]}
\bar{p}_{m,k}[l]= p_{k}[l]   e^{j2\pi lm/M}
\ee
where $p_{k}[l]$ is the channel PDP between the user terminal $k$ and the BS antennas.
 The above equation shows the residual channel that breaks the Nyquist property is characterized by the PDP  response $\tilde{p}_{m,k}[l]$. This effect can be compensated using a prefilter at each subcarrier.  Our proposed prefilter is a fractionally spaced one, \cite{farhang2008signal,farhang2013adaptive}, that, for any $m$, covers the $m$-th band of the filter bank, including the portions of the band that overlaps with the adjacent bands. Therefore, this prefiltering also eliminates the intrinsic interference from the adjacent bands. The prefilter design may be a ZF or an MMSE one that can provide satisfactory performance with minimal taps.  A closer look at equation \eqref{eq:pmk[l]} determines that the prefilter at each subcarrier is the frequency-shifted version of the base-band prefilter and needs to be calculated once for each user.  The proposed prefiltering before precoding, interpolation, and SFB flattens the channel at each subcarrier and improves the output SINR significantly.

\begin{figure*}[t]
		\centering
		\includegraphics[scale=0.1,trim={0 0 0 0},clip]{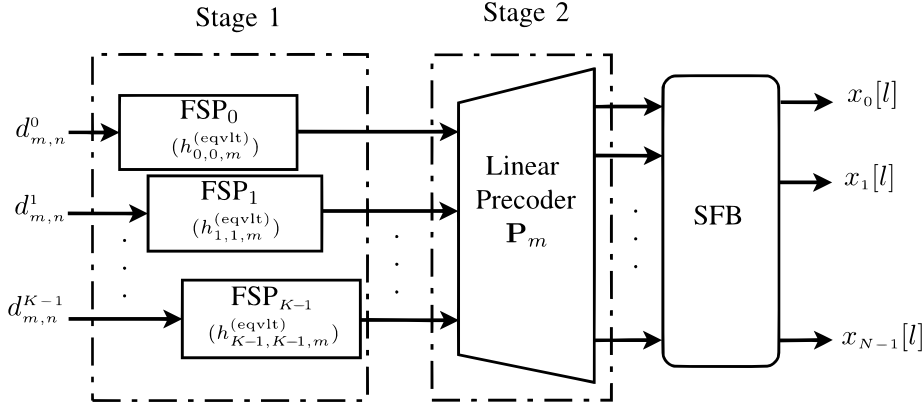}
		\caption{The proposed two-stage precoding scheme. The first stage is a set of FSPs that flattens the channel for the second stage of conventional linear precoding. Conventional linear precoding repeats for each subcarrier $m$ and every user $k$ to remove ISI and ICI.
		}
		\label{blockdiagram}
	\end{figure*}

The proposed two-stage precoding procedure, in the downlink of FBMC-based massive MIMO, is illustrated in Fig.~\ref{blockdiagram}. The data symbols are passed through a set of FSPs followed by a  conventional linear precoder that repeats for each subcarrier $m$ and every user $k$ for removing ISI and ICI.  The first stage can be thought of as a channel flattening step, making it possible for the single-tap precoder to perform optimally. Simulation results that confirm the efficacy of this prefilter design are provided in Section~\ref{sec:simulation}.

 \color{black}
\section{Precoding with Imperfect Reciprocity}
\label{sec:calibration}
In the developments up to this point of the paper, we considered perfect knowledge of the channel at the BS and reciprocal channels in both uplink and downlink directions. However, these assumptions may not be  accurate in a practical scenario as the channel estimates at the BS antennas are obtained during the uplink phase and they may differ from the downlink channel responses. The reason for this is that while the propagation channels may be the same in both directions, different transmitter and receiver chains at the mobile terminals and the BS, that are considered as a part of the channel, are not the same. Moreover, in a  TDD mode, channel aging between the time that channel is estimated and when used for precoding adds additional reciprocity error.
Hence, channel reciprocity errors should considered in designing downlink precoders. 

In the past, a number of channel reciprocity calibration techniques have emerged \cite{kaltenberger2010relative}. However, it has been noted that the reciprocity calibration error can still lead to detrimental effects on the downlink transmission \cite{mi2017massive}. Recently, considering the channel estimation technique in \cite{Hoss2006:Spectrally} and the associated channel estimation error statistics that are provided in \cite{hosseiny2021fbmc}, we have noted that one can quantify the effects of inaccurate channel estimates on signal detection.  According to the results provided in \cite{hosseiny2021fbmc}, the estimation error for the channel tap $l$ between BS antenna $i$ and user $k$, $\Delta h^u_{i,k}[l]$, may be approximated by a complex Gaussian distribution with zero mean and the variance $\sigma_{\rm et}^2 =  \frac{\rm MSE}{K \times L}$. The MSE of the estimation method is also calculated in \cite{hosseiny2021fbmc}. It  has been also noted that the estimation error at a given subcarrier $m$ follows complex Gaussian distribution with the variance $\sigma_{\rm ef}^2=L \sigma_{\rm et}^2.$
That is, $\Delta H_{m}^u({i,k}) \sim \mathcal{CN}(0,\sigma^2_{\rm ef})$.

To tackle the aforementioned issues, in the following, we  start with the channel reciprocity calibration error model, and analyze the calibration and channel estimation error effects on the received signal. We then propose a correction method whose efficacy is confirmed numerically in Section~\ref{sec:simulation}.



\color{black}
\subsection{Reciprocity Calibration Error Model}
Assuming reciprocity calibration to be performed by one of the methods in \cite{kaltenberger2010relative}, RF chains cannot be considered perfectly matched. Here, we extend the narrowband model of \cite{mi2017massive} to FBMC by considering random calibration errors that have a constant gain over each subcarrier band and are independent at different subcarriers. Consequently, these calibration errors can be transformed into the time domain and modeled by a calibration error impulse response at the BS transmit RF chain and any given antenna $i$, i.e. $c_{{\rm t},i}[l]=\mathcal{F}^{-1}\{[\xi_{{\rm t},i}^0e^{j \phi_{{\rm t},i}^{0}}, ..., \xi_{{\rm t},i}^{M-1} e^{j \phi_{{\rm t},i}^{M-1}}] \}$, where $\xi_{{\rm t},i}^{m}$ and $\phi_{{\rm t},i}^m$ denote the magnitude and phase of the calibration error for subcarrier $m$, respectively. Similarly, $c_{{\rm r},i}[l]=\mathcal{F}^{-1}\{[\xi_{{\rm r},i}^0e^{j \phi_{{\rm r},i}^{0}}, ..., \xi_{{\rm r},i}^{M-1} e^{j \phi_{{\rm r},i}^{M-1}}] \}$, $\xi_{{\rm r},i}^m$ and $\phi_{{\rm r},i}^m$ represent the equivalent variables of the receive RF chain for antenna $i$ and subcarrier $m$. Accordingly, uplink and downlink channels can be obtained as 
\be
h^u_{i,k}[l] =  h_{k,i}[l] \star c_{{\rm r},i}[l],
\ee
and
\be
h^d_{k,i}[l] = c_{{\rm t},i} \star h_{k,i}[l],
\ee
respectively. Here, $h_{k,i}[l]$ shows the propagation channel with the PDP defined in Section \ref{sec:flatfbmc}. 

\subsection{Analysis and Compensation}
Here, we present an analysis in the presence of imperfect CSI and reciprocity calibration.  The precoder with imperfect CSI is obtained by substituting $H^d_m({k,i})$ with $\hat{H}^u_m({i,k}) = H^u_m({i,k}) + \Delta H^u_m({i,k})$ in \eqref{eq:combiners}. Therefore, elements of $\bD_m$ in MRT, becomes
\be
\hat{D}_m^{k,k} = \sum_{i=0}^{N-1}|H_m^u({i,k})+\Delta H_m^u({i,k})|^2.
\ee
 Assuming  uncorrelated estimation errors and channel gains, by the law of large numbers, in the asymptotic regime, $\hat{D}_m^{k,k}$ converges to
\begin{align}
 &N \big(\mathds{E}\{|H_m^u({i,k})|^2\}+\mathds{E}\{|\Delta H_m^u({i,k})|^2 \big) \nonumber \\ &~~~= N\big(\mathds{E}\{|H_m({k,i})|^2\}+\mathds{E}\{\xi_{{\rm t},i}^m e^{j\phi_{{\rm t},i}^m} \xi_{{\rm t},i}^m e^{-j\phi_{{\rm t},i}^m}\}+\sigma_{\rm ef}^2\big) \nonumber\\
 &~~~=N+N\sigma^2_{\rm ef},
\end{align}
where, the variance of $\xi_{{\rm t},i}^m$, considered to be small compared to 1, according to models in \cite{mi2017massive}.  \textcolor{black}{Similarly, it can be shown that $(\hat{\bH}_m^u)^{\rm H} \hat{\bH}_m^u$ converges to $\hat\bD_m$. } Therefore, the ZF and MMSE precoders in the asymptotic regime are similar to the MRT. Therefore, in the asymptotic regime, all three precoder converge to $\hat{\bP}_m = \frac{1}{N(1+{\sigma}_{\rm ef}^2)} \hat{\bH}^u_m$.  Consequently, from \eqref{eq:equivalent-channel}, the combined/equivalent channel between the transmit symbol at subcarrier $m'$ of terminal $k'$ and the received one at subcarrier $m$, at the $k$th combiner output converges to 
\begin{align}
        h_{k,k',m}^{\rm (eqvlt)}[l] =&\frac{1}{N(1+{\sigma}_{\rm ef}^2)}\sum_{i=0}^{N-1}  \big(\hat{H}_m({k,i})\big)^* \xi_{{\rm r},i}^m e^{j \phi_{{\rm r},i}^m}  \nonumber \\ 
        &~~~~~~~~~~~~~~~~~~~~~~~\times\big(h_{k',i}[l]\star c_{{\rm t},i}[l]\big). \label{eq:equivalent_channel_IP}
\end{align}
 Moreover, for large values of $N$, \eqref{eq:equivalent_channel_IP}, reduces to
\begin{equation}
h_{k,k',m}^{\rm (eqvlt)}[l] = \frac{1}{1+{\sigma}_{\rm ef}^2}    \mathds{E} \big\{ \xi_{{\rm t},i}^m e^{j \phi_{{\rm t},i}^m} \big(\hat{H}_m({k,i})\big)^*  h_{k',i}[l] \star c_{{\rm t},i}[l] \big\}.
\end{equation}
 Assuming independent channel responses for different users, independent channel taps, and uncorrelated channel estimation errors, one will find that
 \begin{align}
&h_{k,k',m}^{\rm (eqvlt)}[l] = \frac{1}{1+{\sigma}_{\rm ef}^2}    \mathds{E} \big\{ \xi_{{\rm t},i}^m\big\} \mathds{E} \big\{ e^{j \phi_{{\rm t},i}^m}\big\} \nonumber\\&~~~~~~~~~~~~~~~~
\times \mathds{E} \big\{\big(\hat{H}_m({k,i})\big)^*  h_{k',i}[l] \big\}\star \mathds{E} \big\{ c_{{\rm t},i}[l] \big\}.
\end{align}
 Setting $\lambda =  \mathds{E} \{\xi_{{\rm t},i}^m \} \mathds{E} \{ e^{j \phi_{{\rm t},i}^m}\}$, we note that, $\mathds{E} \{c_{{\rm t},i}[l]\}$ is the time domain representation of calibration error and accordingly, converges to an impulse with magnitude of $\lambda$. Additionally, we we have \cite{hosseiny2021fbmc}
\be \label{eq:expectationpdp}
\mathds{E}\big[ \big(\hat{H}_m({k,i})\big)^* h_{k',i}[l] \big] = p_{k}[l] e^{j2\pi lm/M} \delta_{kk'}.
\ee
\color{black}
Thus, the equivalent channel converges to
\be
h_{k,k',m}^{{d}\rm (eqvlt)}[l]  =  \tilde{p}_{m,k}[l]\delta_{kk'},
\label{eq:eq_channel_imperfect_co}
\ee
where 
\be 
\begin{aligned}\label{eq:ColocatedPDP_imperfect}
&\tilde{p}_{m,k}[l]= \frac{\lambda^2 \bar{p}_{m,k}[l]}{1+{\sigma}_{\rm ef}^2}.
\end{aligned}
\ee 

This shows that the effects of channel estimation and reciprocity calibration errors converge to their statistical parameters. It is worth noting that the errors are subcarrier dependent; however, in the asymptotic regime, similar to channel hardening, they average out and converge to an equal value and become frequency independent.  As a result, by modifying our proposed two-stage precoding method, it is possible to compensate for the effects of the reciprocity errors.  

The above results lead to the following conclusion. To compensate for the imperfect CSI and calibration error effect, the scaling factor  $\frac{\lambda^2}{(1+{\sigma}_{\rm ef}^2)}$ should be added to the PDP $\bar{p}_{m,k}[l]$. This is equivalent to adding the {\em correction factor} $\frac{\lambda^2}{(1+{\sigma}_{\rm ef}^2)}$ to the designed prefilter. Since this correction factor may not be known at the BS, as discussed in the next subsection, it may be identified at the receiver/UE through a pilot signal that transmitted by the BS.


\subsection{Pilot-aided Compensation}
The above theoretical development shows that as a result of statistical averaging among the massive MIMO channels between the BS and each UE, the channel estimation and reciprocity errors reduce to a single scaling factor across all the subcarrier bands. This scaling factor may be found at the receiver if the BS transmit a pilot symbol/signal to each UE. The UE uses this pilot to identify the gain factor, i.e., $\frac{\lambda^2}{(1+{\sigma}_{\rm ef}^2)}$, that should be applied to bring the received signal level to the correct value. This clearly is a simple fix that may be simultaneously applied to all UEs in the network, by allocating a single FBMC frame for this purpose, or sending a set of scattered pilots in the payload part of each downlink packet. One may note that pilot symbols of different UEs can overlap in time and/or frequency, since precoder conditions the transmit signals such that each UE only receives its associated information carrying signal. Packets that belong to other UEs are suppressed/nulled.

 \vspace{-1mm} \section{Simulation Results}
 \label{sec:simulation}
 \color{black}
This section is devoted to evaluate our mathematical developments in this paper by computer simulations. We consider a single cell massive MIMO BS with co-located antennas. An SMT system with $M=64$ subcarriers using a PHYDYAS prototype filter, \cite{bellanger2010fbmc}, with overlapping factor $\kappa=4$ is employed to transmit QAM (quadrature amplitude modulation) symbols. Tap delay line-C (TDL-C) from 5G channel model, \cite{etsi2017138}, used to generate channels.  This model provides a PDP based on a normalized root mean square (RMS) delay spread. The normalized RMS delay spread is randomly scaled for different users in each simulation instance is scaled in the range of  $[90~{\rm ns},110~{\rm ns}]$, i.e., for channels with moderate lengths,  \cite{etsi2017138}.  This is due to non-equal PDPs between the users and the BS antennas in practical scenarios.  Additionally, a normalized PDP is assumed, i.e., $\sum_{l=0}^{L-1}p_k[l]=1$ for $k=0,\ldots, K-1$. \color{black} We consider average transmit power of $K$ from BS antennas which distribute equally between different users, accordingly, we set signal-to-noise (SNR) of $10$~dB at each user terminal. Sampling frequency of $15.36$~MHz considered in this experiment which results in the subcarrier spacing of $240$~kHz and is inline with 5G NR specifications, \cite{etsi2017138}. To model the magnitude of reciprocity calibration error, we consider $\xi_{{\rm t},i}^m$ and $\xi_{{\rm r},i}^m$ with uniform distribution between $0.98$ and $1.02$.  Moreover, $\phi_{{\rm t},i}^m$ and $\phi_{{\rm r},i}^m$ considered to have uniform distribution between $-\frac{2\pi}{9}$ and $\frac{2\pi}{9}$ following measurements in \cite{3GPPmeasurement}. We have obtained our results for 1000 independent realizations of the channel with $K=4$ users. 
 \color{black}

\begin{figure}[t]
		\centering
		\vspace{-2mm}
		\includegraphics[scale=0.65,trim={0 0 0 0},clip]{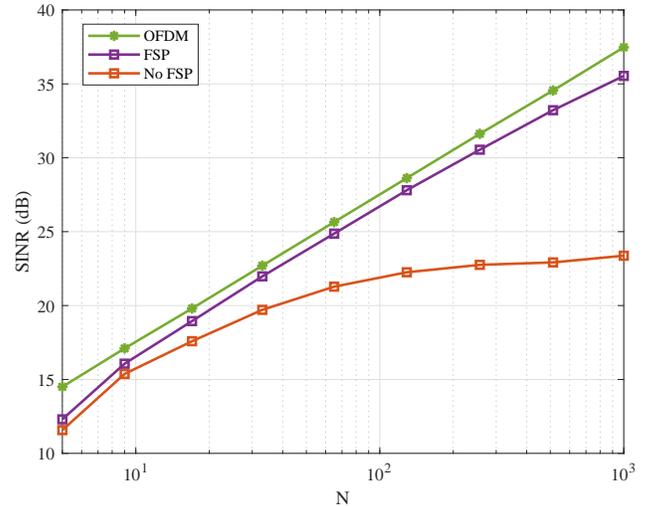}
		\caption{SINR vs. the number of BS antennas, N. The FSP design is based on the PDP of the underlying channels, $L_{\rm FSP}=5$ is considered.}
		\vspace{-2mm}
		\label{fig:SINRperfect}
	\end{figure}

In Fig.~\ref{fig:SINRperfect}, we evaluate the performance of our proposed FSP, measured by signal-to-interference-plus-noise ratio (SINR). The FSP design follows the same procedure as fractionally spaced equalizer (FSE) design proposed in  \cite{hosseiny2021fbmc}, for uplink. Here, in the designs the FSP length is set equal to $5$.  In this experiment, we assumed perfect knowledge of channel.  These results show that the proposed FSP significantly improves the performance. In addition, in Fig.~\ref{fig:SINRperfect} the result of an OFDM precoder are presented for a bench mark. We note that, by construction, when cyclic prefix (CP) length in OFDM is sufficiently, subcarriers are subject a set of perfect flat gains, hence, with an ideal precoding FBMC performance should match that of OFDM. The results presented in Fig.~\ref{fig:SINRperfect} confirms a near perfect operation of the two-stage precoder that is presented in Fig.~1.

\begin{figure}[t]
		\centering
		\vspace{-2mm}
		\includegraphics[scale=0.65,trim={0 0 0 0},clip]{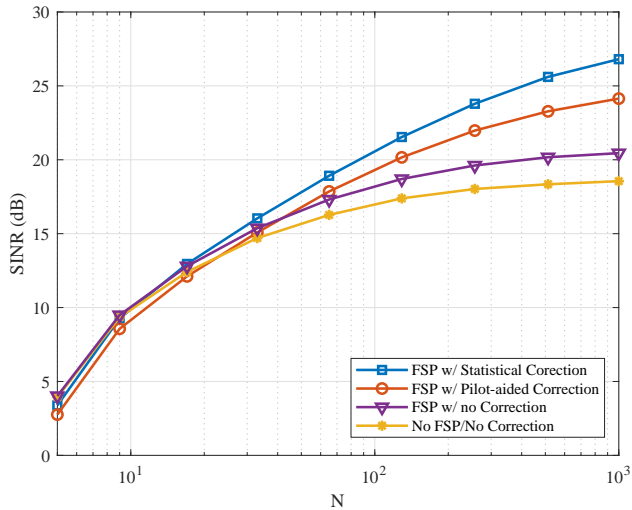}
		\caption{Output SINR vs. the number of BS antennas, N, for different FSP
designs with $L_{\rm FSP} = 5$.}
		\label{fig:SINRimperfect}
	\end{figure}

To evaluate our proposed design in a practical scenario, we deploy the channel estimation method of \cite{Hoss2006:Spectrally}, performed in uplink. Also, to obtain the scaling factor for compensation of calibration and channel estimation errors, a set of pilots are transmitted of the first FBMC block of each downlink packet.  The results of this study are presented in Fig.~\ref{fig:SINRimperfect}. As seen, with no compensation, FSP provide a small  gain of about 2~dB for $N>100$. This gain increases to 5 to 10~dB if the compensation scaling factor is known perfectly. For our simulation setup, estimation of the scaling factor through pilots incurs 1  to 3~dB loss in performance. However, we believe with better design of pilots this gap may be reduced. Some research into the details of pilot designs for this purpose is underway.

 \vspace{-1mm}\section{Conclusion}
\label{sec:conclusion}
In this work, we developed a practical precoding method for the downlink of FBMC-based massive MIMO in a co-located antenna setup. Theoretical results that show the impact of channel estimations error and reciprocity mismatch in uplink and downlink radio chains were developed. The proposed method includes a two-stage precoder. The first stage of the precoder applies a fractionally spaced prefilter (FSP) for flattening/equalizing the channel across each subcarrier band. The second stage is the conventional precoder, like the maximum ratio transmission (MRT), the zero forcing (ZF), or the minimum mean square error (MMSE),   whose goal is to separate different users' data symbols. We also studied the theoretical impact of calibration and channel estimation errors and showed that in massive MIMO, these errors can trivially be obtained by sending a pilot signal and compensated through a single scaling factor that is similar for all the subcarriers, thanks to the channel hardening effect in massive MIMO systems. Simulation results that corroborate our theoretical findings were also presented. 
\vspace{-2mm}

\ifCLASSOPTIONcaptionsoff
  \newpage
\fi

\bibliography{ref.bib}
\bibliographystyle{IEEEtran}
%


\end{document}